\documentclass[conference]{IEEEtran}
\input epsf
\usepackage{graphicx}
\begin{document}

\title{\LARGE Applying Cognitive Tutoring in the use of Bioinformatics Tools}

\author{\authorblockN{Angela Makolo, Daniel Nkemelu}
\authorblockA{University of Ibadan, Nigeria \\
aumakolo@gmail.com, daniel.nkemelu@gmail.com}}



%


\maketitle

\begin{abstract}
With the proliferation of simple and complex bioinformatics tools, there is the need to teach researchers how to use these tools effectively. To evaluate the potential of cognitive tutoring in the wide-scale adoption of several bioinformatics tools, we designed a simple prototype. We embedded a cognitive tutor, built with the Cognitive Tutor Authoring Tool, on a pre-existing platform, the Gene Adjacency Program, developed by the University of Ibadan Bioinformatics group. Our preliminary tests  show that researchers who used the platform with the cognitive tutor embedded showed higher levels of competence and efficiency. These results indicate that cognitive tutors have the potential to teach bioinformatics researchers employing new tools how to efficiently use them and accurately make sense of their results.
\end{abstract}
\IEEEoverridecommandlockouts

%
\IEEEpeerreviewmaketitle

\section{Introduction}
A major activity in bioinformatics is to develop software tools to generate useful biological knowledge. There are several bioinformatics tools created with the goal of developing methods for the analysis and interpretation of complex datasets. Bioinformaticians make use of these tools to understand and discover novel solutions to the vast amount of biological problems available. The challenge, however, is on how to teach researchers, especially those with non-technical backgrounds, to make effective use of these tools.

Tutoring done by humans has proven to be a very effective form of instruction and there is evidence that expert tutors produce enormous learning gains \cite{chi}. However, with recent advancements in cognitive computing and artificial intelligence, computers have become efficient, flexible, scalable and economic alternatives to expert human tutors \cite{chen}.

Human tutoring is widely believed to be the most effective form of instruction and tutoring available, and both experimental work and historical evidence confirm that expert human tutors can produce extremely large learning gains. Ever since computers were invented, they seemed capable of becoming both collaborators and alternatives to human experts. In recent years, cognitive tutors have been employed to help tutor people and the gains have been largely significant \cite{vanlehn}.

According to \cite{alaven}, a cognitive tutor is a particular kind of intelligent tutoring system that utilizes a cognitive model to provide feedback to users of complex systems as they are working through problems. The Cognitive Tutor programs utilize cognitive models and are based on model tracing and knowledge tracing. Most tutors perform user evaluation by flexibly comparison of their activity against generalized examples of problem-solving actions. Such tutors are referred to as example-tracing tutors. Example-tracing tutors are capable of sophisticated tutoring behaviors; providing step-by-step guidance on complex problems while recognizing multiple user strategies and, where needed, maintaining multiple interpretations of user behavior.

As more bioinformatics tools are developed to solve biological problems, researchers are faced with the challenge of learning how to use these tools and accurately making
sense of their results. The problems faced include: inability to use the tools, inability to effectively use the tools to get proper result and inability to make sense out of its result. The implementation of a cognitive tutor for bioinformatics tools is very important because it helps researchers without technical skills to efficiently use bioinformatics tools thereby helping to eliminate or at least reduce errors resulting from improper use of the tools.

In this work, we developed a cognitive tutor for a bioinformatics tool; the Microbial Gene Adjacency Visualization Software \cite{makolo}. The inputs processed by the application program were RefSeq files downloaded from \cite{refseq}. 
A web interface was designed to host the tutor. The cognitive model on which the tutor runs was built using the Cognitive Tutor Authoring Tool$^{\tiny{\textregistered}}$ provided by Carnegie Learning \cite{blessing}; by generalizing and annotating the behaviour graph and creating the Behaviour Recorder file [.brd] using extensible markup language (XML), correct and incorrect behaviours were demonstrated.

Results from the new system show that using the cognitive tutoring approach to teach bioinformatics can increase widespread adoption of tools and further progress in the field.
\begin{figure}[ht]
\includegraphics[width=0.5\textwidth]{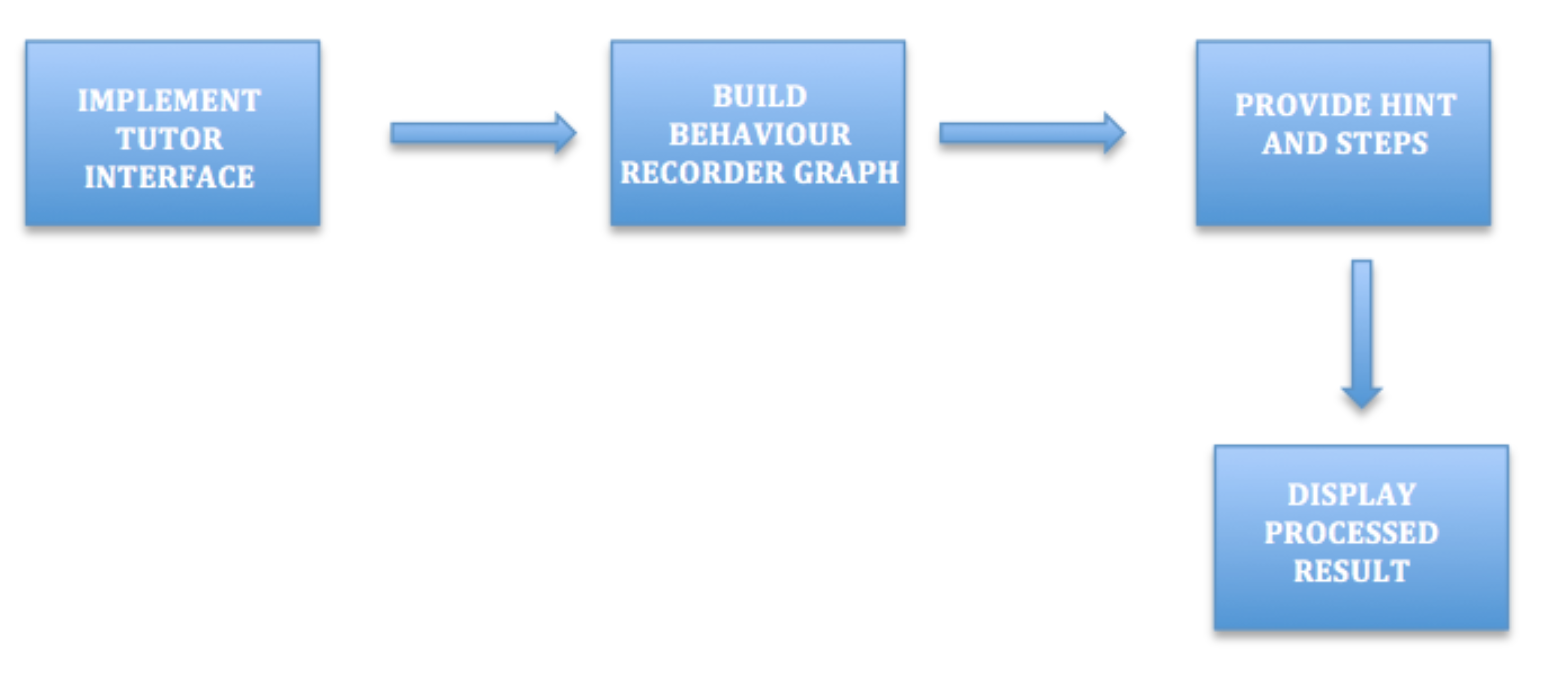}
\caption{ Steps in setting up a cognitive tutor on an interface using the Cognitive Tutor Authoring Tool$^{\tiny{\textregistered}}$.}
\end{figure}

\section{Background Study}
\cite{chiew} defined cognitive computing in terms of cognitive
informatics, as a multidisciplinary field that applies how the brain processes information and copes with decision making. The focus of cognitive computing is on mimicking the mechanisms of the brain to endow computer systems with the faculties of feeling, thinking and knowing such as is evident in intelligent tutoring systems and cognitive tutors \cite{octavio}. In the following subsections, we will review existing literatures in the subject matter and introduce concepts in our overall work.

\subsection{ Intelligent Tutoring Systems (ITS) }
An Intelligent Tutoring System (ITS) is a computer system that has the objective of providing immediate and customized instruction or feedback to learners, usually without
intervention from a human teacher \cite{psotka}. ITSs have the common goal of enabling learning in a meaningful and effective manner by using a variety of computing technologies. According to \cite{psotka}, ITSs are being used in both formal education and professional settings and have demonstrated their capabilities and limitations. Modern day ITS tries to replicate the role of a teacher or a teaching assistant and involves problem generation, intelligent automatic feedback generation with a high recall value.

\subsection{ Cognitive Tutor }
Anderson et al \cite{anderson}, defines a Cognitive Tutor as a kind of intelligent tutoring system that utilizes a cognitive model to provide feedback to users as they are working through problems. This feedback immediately informs users of the correctness, or incorrectness, of their actions in the tutor interface. In addition, it also has the ability to provide context-sensitive hints and instruction to guide users toward reasonable next steps. According to them, Cognitive Tutors were
originally developed to test ACT-R theory for research purposes and they are developed also for other areas and subjects such as computer programming and science. 

The Cognitive Tutor programs are based on model tracing and knowledge tracing. Model tracing and knowledge tracing are essentially used to monitor students' learning progress, guide them to the correct path to problem solving, and provide feedback by checking actions like button clicks, and value entering. This enables learners to develop a complex problem-solving skill through practice. Typically, cognitive tutors provide such forms of support as: (a) a problem-solving environment that is well designed and allows for "visible thinking" (b) step-by-step feedback on student performance (c) feedback messages specific to errors (d) context-specific next-step learning hints at student's request, and (e) individualized problem selection. Cognitive Tutors accomplish two of the principal tasks characteristic of human tutoring:
\begin{enumerate} 
 \setlength{\itemsep}{-2ex}  
 \setlength{\parskip}{0ex} 
 \setlength{\parsep}{0ex}
\item monitors the student's performance and provides context specific individual instruction, and.\hfil\break
\item monitors the student's learning and selects appropriate problem solving and learning activities.\hfil\break
\end{enumerate}

\subsection{The Cognitive Tutur Authoring Tool (CTAT$^{\tiny{\textregistered}}$)}
Cognitive Tutors and Example-Tracing Tutors, the two types of tutors supported by CTAT$^{\tiny{\textregistered}}$, represent different trade-offs between ease of authoring on the one hand and generality and flexibility of the resulting tutors on the other. Cognitive Tutors are rooted in the ACT-R theory of cognition and learning. They are capable of interpreting users problem-solving behavior by employing a cognitive model that captures, in the form of production rules, the skills that the user is expected to learn \cite{alaven}.
Authoring a Tutor involves the following development steps: (a) Create the graphical user interface (GUI) used by the student, (b) Demonstrate alternative correct and incorrect solutions, (c) Annotate solutions steps in the resulting “behavior graph” with hint messages, (d) feedback messages, and labels for the associated concepts or skills, (e) Inspect skill matrix and revise.

CTAT’s Example-Tracing Engine uses the Behavior Graph to guide a user through a problem, comparing the student’s problem-solving behavior against the graph. It provides positive feedback when the user’s behavior matches steps in the graph, and negative feedback otherwise \cite{koedinger}.

\subsection{Cognitive Tutoring in Bioinformatics and the Gene Adjacency Program}
Bioinformatics can be described as a discipline that integrates computers, software tools, and databases in order to address biological questions. According to \cite{makolo}, visualization models are critical to understanding and making sense of big and complex data generated from genomic research. The Gene Adjacency Program is a model for the visualization of neighborhood genes and their representation as binary codes. This concept of using binary code for modeling is derived from computational thinking techniques which simulates problems using computer logic of applying abstraction and pattern matching to extract hidden patterns aimed at knowledge discovery. The binary representation enables easy pattern matching of the different gene component
and the comparative analysis of multiple genomes and prediction of transcriptional units which are the basis of biomolecular network or biosynthetic pathways. We will embed a cognitive tutor to this tool for our present research.

At present, the most striking attempt at applying cognitive tutoring in bioinformatics is the CTAT Genetics which contained a total of 12 lessons developed for the SimBioSys genetics tutor project at CMU, developed primarily by Albert Corbett, Benjamin MacLaren and Linda Kauffman. 

\subsection{Limitations of the Cognitive Tutor}
Despite the commercial successes of Cognitive Tutor, \cite{valeven} identified a design limitation associated with it. Its complexity demands that designers spend 100 of hours per instructional hour to create the program. Furthermore, Cognitive Tutors may not account for the flexible, complex and diverse ways humans create knowledge as cognitive model is based on assumptions about how learning occurs which dictates the chosen instructional methods such as hints, directions and timing of the tutoring prompts. Thus, human tutors may outperform Cognitive Tutor if it provides a higher level of responsiveness to student errors as they are capable of providing more effective feedback and scaffolding to learners than Cognitive Tutors. This indicates that the cognitive model may still be incomplete but has the potential for massive improvement \cite{scandura}.

\section{System Design}
Most users of bioinformatics tools depend on human tutors and/or technical documentation for understanding the flow and how to use the tools. However, with the proliferation of bioinformatics tools \cite{kanehisa}, human tutors are no longer sufficient to properly tutor the large number of researchers and biologists who are interested in using the tools for research and other breakthrough findings. Another challenge with human tutors is the idea that they may not fully cover the overall features of the tool at every teaching instance and may not be readily available at the convenience of the user. The problem of miscommunication by human tutors is also a difficulty \cite{michael}.

The proposed system involves the implementation of a cognitive tutor on the Gene Adjacency Program \cite{makolo}. The implementation of a cognitive tutor for bioinformatics tools is very important because it ensures learning by doing as hints are provided based on current learning challenge and next steps. It would also help researchers without technical skills to efficiently use the program, thereby eliminating or at least significantly reducing errors resulting from improper use of the program.

The design objectives include:
\begin{itemize} 
\item {• Designing a host user interface for the tutor in the bioinformatics application.}
\item {• Configuring the Cognitive Tutor that would determine the flow of tutoring.}
\item {• Setting up the cognitive tutor on the selected bioinformatics tool.}
\item {• Providing clear and unambiguous hints to users where necessary.}
\item {• Testing the application with real data and users to ensure that it guides users to carry out the primary function of the host tool.}
\end{itemize}

\subsection{Software Requirements}
The Cognitive Tutor Authoring Tool$^{\tiny{\textregistered}}$ is platform independent, hence it can be configured on Mac OSX, Windows and Linux distributions. MAMP Server (Mac OSX, Apache, MySQL and PHP Server) was used on Mac OSX to process all requests and display results. The system was developed using HTML5 (Hypertext Markup Language) – for building the web interface, JavaScript – for embedding the cognitive tutor, Object Oriented PHP – for coding the system logic and XML (Extensible Markup Language) – for extending the Behaviour Recorder File (.brd) needed by the cognitive tutor.

\subsection{How to Use the Program}

\begin{figure}[ht]
\includegraphics[width=0.5\textwidth]{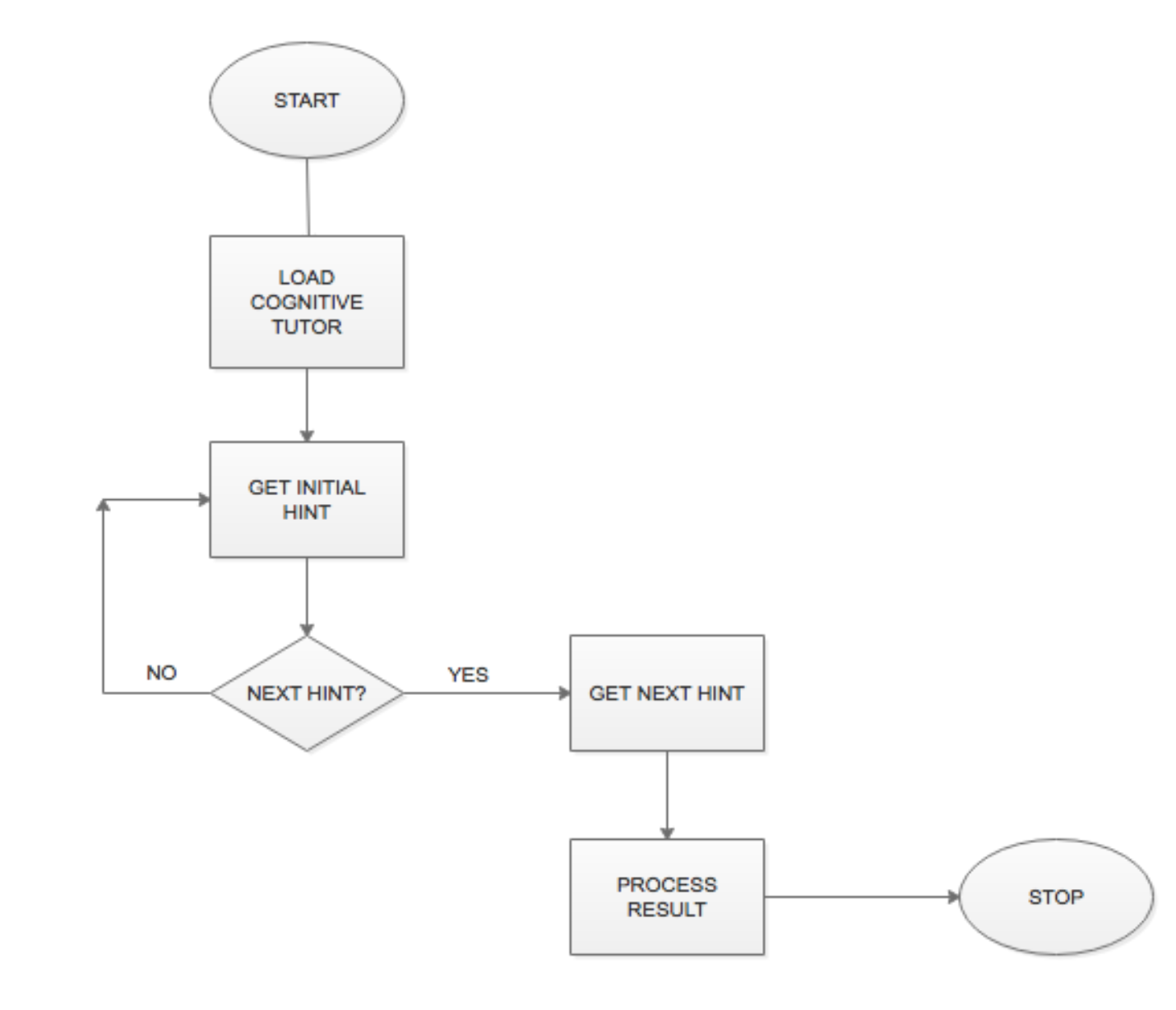}
\caption{ Flow diagram showing how to use the cognitive tutor.}
\end{figure}

\begin{enumerate} 
 \setlength{\itemsep}{-2ex}  
 \setlength{\parskip}{0ex} 
 \setlength{\parsep}{0ex}
\item {\itshape Load the application which activates the cognitive tutor}\hfil\break
\item {\itshape Click on the HINT button to get the first instruction from the tutor.}\hfil\break
\item {\itshape Follow the instruction by clicking on CHOOSE FILE button of the Gene Adjacency Program and selecting your downloaded RefSeq file.}\hfil\break
\item {\itshape After selecting the file, click on NEXT pointer to get next instructions. At this point, the PREV pointer is activated to allow user go over previous steps.}\hfil\break
\item {\itshape Click on the PROCESS FILES button to view and download processing result.}\hfil\break
\item {\itshape Click DONE to complete the tutoring process and get your results.}
\end{enumerate}

\begin{figure}[ht]
\includegraphics[width=0.5\textwidth]{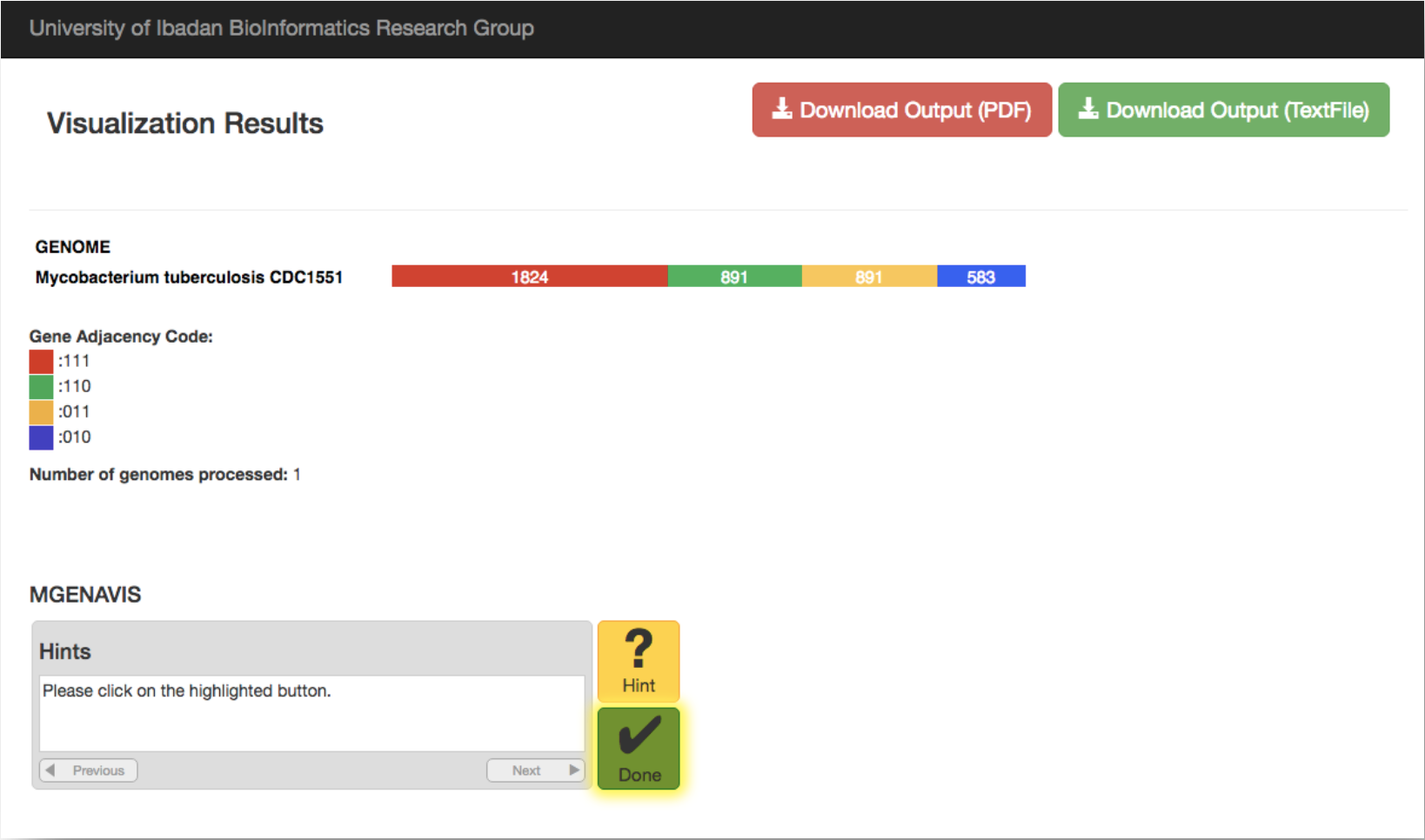}
\caption{ User interface of the Gene Adjacency Program showing the cognitive tutor in the bottom left.}
\end{figure}

\begin{figure}[ht]
\includegraphics[width=0.5\textwidth]{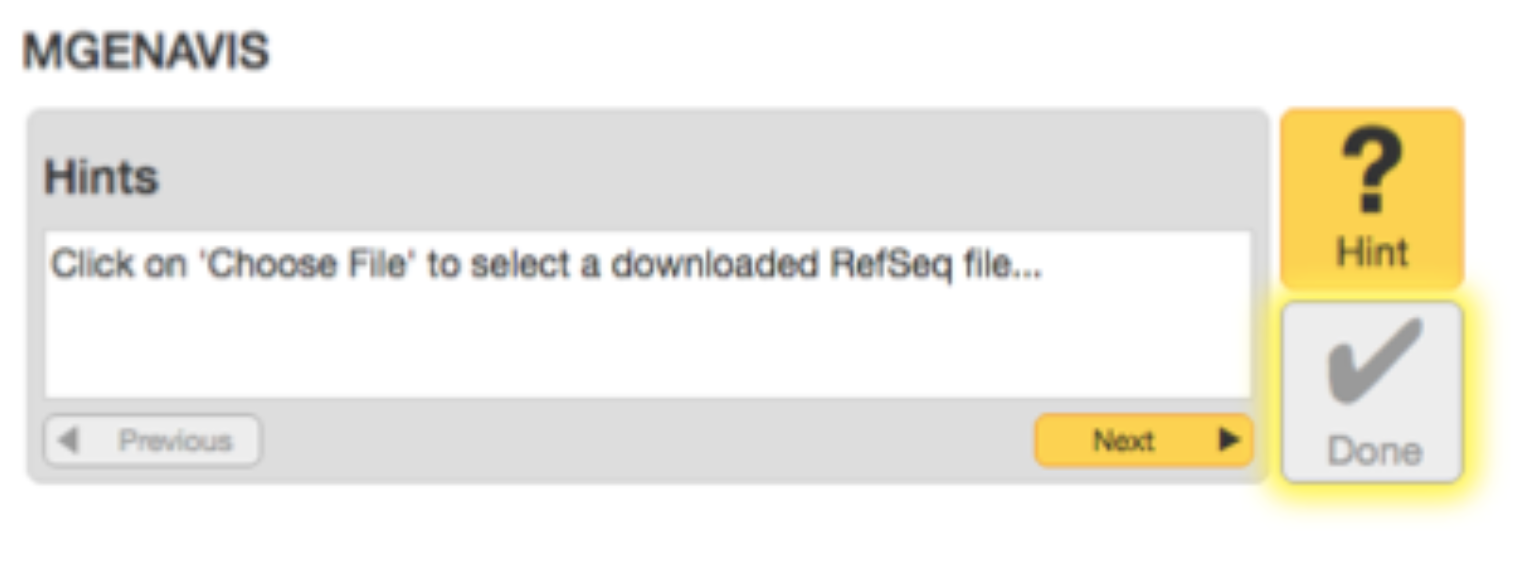}
\caption{ The cognitive tutor displaying learning instructions to the user.}
\end{figure}

\section{Results and Discussion}
On launching the application, the first interface displays the cognitive tutor. At this point, it loads the logic in the Behaviour Recorder File onto the index web interface of the Gene Adjacency Program. After successfully loading the tutor, the homepage displays a brief textual explanation of
what the program is about and an instruction to follow the Cognitive Tutor for directions on how to use the program. The user clicks on the HINT button to get the first learning instruction from the tutor. The user then follows the instruction and clicks on the CHOOSE FILE button of the Gene Adjacency Program and selects the file. It is expected that the user has downloaded a RefSeq file with the extension RefSeq.cds.tab that will be processed by the application from \cite{refseq}. 

After selecting the desired file, click on the NEXT button on the cognitive tutor to get the next learning instruction. If the user wishes to process more than one RefSeq file, click on the ADD FILE button and repeat Step 3 to add a new file for processing, else, the user can click on the NEXT button to proceed. Notice that the PREVIOUS button is activated to allow the user go over previous steps either to correct mistakes or to confirm mastery of the steps. Click the NEXT button for more learning hints. Then click on the PROCESS FILES button to view processing result. Notice that you can download result in PDF or txt formats. Click on the DONE button to complete the tutoring session.

\paragraph{Preliminary Evaluation}
To preliminarily evaluate our system, we carry out a small user testing experiment with 6 bioinformatics researchers having varying experience levels and all having no prior experience using the Gene Adjacency Program. The group was divided into 2 subgroups: test and control. Each member of the test group was given the application with the cognitive tutor embedded while each member of the control group was given the application without the cognitive tutor.

By observation and oral feedback, we find that the test group showed higher levels of competence and efficiency using the system without the need of an expert tutor compared to the control group. While the control group users needed human help to navigate the program and took longer times to get results, the test group were satisfied with the system and confirmed that the cognitive tutor improved their ability to use the Gene Adjacency Program; and that the learning instructions were simple and direct, allowing them navigate through the application with ease.

\paragraph{Future Work}
The result of this project is of immense use to bioinformaticians and biologists and it has
a large capability for expansion. Further expansions could be made such as:
\begin{itemize} 
\item{Using the cognitive model instead of example-tracing models in order to cover a much broader scope of bioinformatics tools and use cases.}
\item{Data analysis of user performance and experience should be made in order to appropriately and statistically quantify the overall impact of the system.}
\end{itemize} 

\section{Conclusion}
In this work, we focused on exploring the concept of cognitive tutoring and how it can be applied to better use and understand bioinformatics tools. We showed that cognitive tutors, which have been previously applied in other fields such
as Mathematics, Economics and Programming, could be applied in Bioinformatics learning and research with
considerably large gains. We implemented this by applying a cognitive tutor to the Gene Adjacency Program developed by the University of Ibadan Bioinformatics Group and demonstrating how tutors can be useful in helping non-technical users adapt to new bioinformatics tools and technology by providing intelligent hints and learning instructions.


\end{document}